IoT Security: On-Chip Secure Deletion Scheme using ECC Modulation in IoT Appliances

Na Young Ahn [1] and Dong Hoon Lee [2]

[1] Institute of Cyber Security & Privacy, Korea University, Seoul, South Korea (humble@korea.ac.kr)
[2] Graduate School of Information Security, Korea University, Seoul, South Korea (donghlee@korea.ac.kr)

*Abstract*— NAND flash memory-based IoT devices inherently suffer from data retention issues. In IoT security, these retention issues are significant and require a robust solution for secure deletion. Secure deletion methods can be categorized into off-chip and on-chip schemes. Off-chip secure deletion schemes, based on block-level erasure operations, are unable to perform real-time trim operations. Consequently, they are vulnerable to hacking threats. On the other hand, on-chip secure deletion schemes enable real-time trim operations by performing deletion on a page-by-page basis. However, the on-chip scheme introduces a challenge of program disturbance for neighboring page data. The proposed on-chip deletion scheme tackles this problem by utilizing ECC code modulation through a partial program operation. This approach significantly reduces the program disturbance issue associated with neighboring page data. Moreover, the proposed code modulation secure deletion scheme allows for real-time verification of the deletion of original data.

*Index Terms*—Deletion, NAND Flash Memory, ECC Modulation, Verification, Partial Program, Program Disturbance

1. Instruction

In the future, the legal enforcement of the duty to delete data on NAND flash memory is highly likely [1, 2, 3]. Specifically, anti-forensics techniques for NAND flash memory play a crucial role in safeguarding personal information stored not only in offline electronic devices but also in numerous online-connected devices. Many electronic devices utilize storage devices based on NAND flash memory, which can either be embedded or externally connected. These electronic devices are interconnected through various networks. The emerging threat of connecting these electronic devices to high-performance computers, such as commercialized quantum computers, poses risks. Malicious hackers could potentially access the personal information of multiple users at any given time. Even if the collected personal information is encrypted, hackers with quantum computing capabilities would be able to decrypt the encrypted data with ease [4, 5].

Generally, NAND flash memory is inherently vulnerable to digital forensics [6,7]. Write Amplification Factor (WAF), one of the performance indicators of NAND flash memory, is larger than 1 [8-11]. Many studies are being conducted to lower WAF, such as efficient garbage collection and temperature-aware data storage space distinction, but there are limitations. Fundamentally, the original data stored in NAND flash memory contains not only original data stored in a space visible to legitimate users, but also original data stored in a space not accessible to legitimate users [12]. For spaces inaccessible to legitimate users, for example, over-provisioning (OP) areas, forensics issues may be caused by malicious users. Secure deletion (or data sanitization) using TRIM techniques and encryption techniques has been introduced [13-16]. However, these techniques are still exposed to forensics threats because they are not real-time secure deletion techniques in terms of performance [17-19]. In fact, studies on forensic results on used phones and used storage devices have been introduced by several researchers. Recently, anti-forensics techniques have been introduced for these forensic issues[12,20-27]. In particular, techniques for performing secure deletion in real time without performance degradation have been introduced. Real-time secure deletion techniques are emerging as a major anti-forensics technology. At the same time, it is necessary to introduce a verification technique of such real-time secure deletion. This paper introduces a verification technique of real-time secure deletion.

Recently, there has been a debate regarding the possibility of a malicious code injection attack on the invalidation area, particularly the OP (Over-Provisioning) area [26]. These problems arise due to the incomplete nature of the existing trim operation. These issues are inevitable because the trim operation does not perform real-time trimming. Although the host sends a trim command, the storage device only simulates the trim operation instead of actually executing it, resulting in the retention of the original data in the storage device. The page unit secure deletion technique offers the advantage of performing real-time trim operations.

In Section 2, we will review existing secure deletion schemes, such as off-chip secure deletion schemes. In Section 3, we discuss the limitations of current secure deletion technologies. In Section 4, we propose the ECC (Error Correction Code) modulation secure deletion scheme. Section 5 presents a performance comparison between existing on-chip secure deletion schemes and the proposed on-chip secure deletion scheme. This paper aims to propose an on-chip secure deletion technique that is cost-effective, reasonable, and easily implementable for real-time trim operations. Further research and studies must be conducted before applying this technique to actual products, and it is expected that these studies will be actively pursued in the near future.

2. Related Works

When the operating system deletes a file, it doesn't actually delete the file itself, but rather removes the metadata associated with the file. File systems manage both the actual files and metadata, which contains essential information for file management. This metadata includes details such as the file's storage location, size, name, and access permissions [28]. Conversely, if the metadata for a specific file is lost, the file is considered non-existent within the storage device. In other words, deleting the metadata also implies deleting the file itself. Secure deletion involves the complete removal of both the file's metadata and the actual file.

2.1. TRIM command

The TRIM command refers to the actual deletion of a file from the storage device after its metadata has been removed [29]. TRIM is performed to completely delete invalidated data within the storage device and free up necessary space. Enabling TRIM settings can enhance the write speed of the storage device, optimize space utilization, and extend its lifespan. The auto TRIM function is supported in the Windows system, and the storage device internally supports TRIM. Typically, the TRIM command is sent from the host to the storage device. The TRIM command is an important command used in Solid State Drives (SSDs). When a file system deletes a file, the operating system simply marks that space as 'available', rather than removing the actual data. However, because SSDs write and delete data in block units, they need to completely erase the previous data before writing new data to the 'available' space. This process can lead to slowdowns. The TRIM command solves this problem. It informs the operating system that a file has actually been deleted, allowing the SSD to clean up that space in advance. This significantly reduces slowdown when writing new data.

2.2. Secure deletion schemes

One of the secure deletion technologies performed in SSDs involves using the TRIM command. The TRIM command notifies the operating system that a file has been deleted, and through this, the SSD cleans up the space and prepares to write new data. However, this alone does not completely delete the data but only makes it writable, making data recovery possible. To compensate for this, the 'Secure Erase' function is used. Secure Erase is a command that completely initializes all cells of the SSD, making data recovery impossible. This function is especially important when permanently deleting sensitive information and is one of the most effective ways to completely erase data from an SSD. The combination of these two functions is crucial for securely managing data on SSDs.

2.2.1. Erase operation

The read/program operation and erase operation sizes in NAND flash memory are determined based on the physical structure [20]. NAND flash memory consists of multiple memory blocks, each containing several pages, and each page contains multiple memory cells connected to a wordline. Generally, read/program operations are performed on a page size, while erase operations are performed on a block size. To manage the lifespan of NAND flash memory, erase operations are kept relatively minimal. This implies that there may be a time difference between when the user intends to delete a file and when the file is completely erased from the storage device due to these inherent limitations. Erase operations are performed block by block, removing charges in the charge trap layer. In contrast, program/read operations are conducted on pages, which are significantly smaller in size compared to blocks. However, the existing TRIM operation is based on erase operations. Thus, while TRIM is crucial for security purposes, it can impact performance and lifespan management of the storage device. The TRIM operation typically involves garbage collection before the erase operation [30]. Garbage collection refers to the process of collecting valid pages from at least two or more blocks and moving them to a new block. Only the valid pages from multiple blocks are copied to a new memory block during garbage collection. As a result, a new block is composed of valid pages accessible by the host, while the original blocks are regenerated as reusable memory blocks through an erase operation. Since garbage collection can negatively affect storage device performance, it is usually performed as a background operation or during idle states.

2.2.2. Encryption key deletion

Data is encrypted and stored using an encryption key, and when data is deleted, only the encryption key is removed [30]. This encryption scheme allows for fast and secure deletion. However, deleting all keys in a file can introduce significant overhead. As an alternative, a proposed scheme suggests storing all keys in one memory block and deleting them together. However, this encryption key deletion scheme may not be easily applicable to a real-time invalidation scheme used in databases.

### 2.2.3. Scrubbing

To achieve undetectable secure deletion, scrubbing involves removing deleted data, making it inaccessible to the host, and hiding the deletion history to prevent knowledge of past deletions. The full scrubbing scheme ensures that all page data is overwritten with zero pages [10, 20]. The partial scrubbing scheme involves making a portion of page data zero. One example of partial scrubbing is NAND Flash Partial Scrubbing (NFPS). This scheme makes it challenging to detect the existence of deleted files, and partial scrubbing may include partial page reprogramming.

### 2.2.4. Deletion pulse application

The deletion pulse application scheme is similar to the page unit reprogramming scheme [12, 23]. However, while the reprogramming scheme involves programming data with specific values, such as zero bits, the deletion pulse application scheme applies multiple deletion pulses to the corresponding wordline to perform partial overwriting, even if the most significant bit is not necessarily obtained. The level and number of deletion pulses can be predetermined, such as changing the data to an extent that it cannot be recovered by the error correction code. Through the deletion pulse application scheme, the original data is altered to arbitrary data. This scheme can be particularly useful and easily applicable in the case of multi-bit memory cells.

### 3. Incomplete TRIM

The trim operation essentially refers to the complete deletion of a file from a storage device based on the host's request. When a trim-related command is sent to the storage device, the host assumes that the operation is completed upon receiving a completion message indicating the trim operation's execution. However, in reality, this completion is only internally handled by the storage device and goes unrecognized by the host. Consequently, it is not uncommon for only the associated metadata to be deleted without actually removing the original data. Unfortunately, the majority of secure deletion schemes currently suffer from this limitation, making forensic analysis of storage devices relatively easy. This is primarily because the original data remains intact while only the related metadata is deleted, leading the host to believe that secure deletion is accomplished. Hackers are actively seeking ways to recover metadata, while storage device administrators are exploring strategies to make metadata recovery impossible. From the customer's point of view, a full trim operation will be required. However, in the existing secure deletion schemes, when performing a complete trim operation, an erase operation must be performed on a valid block or an invalid block to delete original data. Since these schemes are unreasonable in terms of the lifespan and cost of the storage device, manufacturers of the storage device are proceeding with secure deletion at a level that partially compromises the request of the host. In the end, the customer is requesting a full trim operation, but the storage device does not 100% accept this request, and is performing an incomplete trim operation in the middle of the line.

This problem primarily arises from performing erase operations at the block level and program operations at the page level, a well-known fact among experts in the field. However, despite their awareness of the solution involving deletion operations at the page level, experts have not pursued it extensively. Some researchers have suggested this solution as early as 2016 [13-15]. For instance, a scrubbing scheme was presented in 2016, followed by an arbitrary data overwriting scheme and a deletion pulse application scheme in 2017 [12]. In 2018, a down-level program scheme was proposed to modulate a multi-level program into a single-level program [21]. We refer to these page-level secure deletion schemes as on-chip secure deletion schemes, while other schemes are commonly known as off-chip security schemes.

### 3.1. Limitations of existing schemes

Off-chip secure deletion schemes have inherent limitations. They instruct the host to perform a trim operation and falsely report completion. While this may seem to have the same effect as actual deletion based on the strict definition of trim, it is incorrect. Experts in the field understand that this claim is flawed. For instance, if the original data is encrypted and the encryption key is deleted, it is argued that the original data is unlikely to be recovered without the encryption key. However, technology is advancing rapidly, with the imminent commercialization of quantum computing. This means that even hackers can utilize quantum computing technology, making the possibility of recovering encrypted data undeniable. If the cost of recovering encrypted data is lower than the cost of using quantum computing, hackers will undoubtedly attempt to recover the encrypted original data. The management of storage devices is further complicated by the WAF (Write Amplification Factor). Due to the different erase units and program units within a storage device, the WAF increases over time. In general, storage devices contain invalidation blocks and validation blocks. As the WAF increases, the occurrence of these invalidation and validation blocks becomes more frequent. However, invalidation blocks contain only invalidated pages, whereas valid and invalid blocks may contain both invalidated and valid pages. This situation makes it challenging to manage the original data. One or more instances of original data may exist in validation blocks, and even in invalidation blocks. When the host issues a command to delete the original data in the storage device, a full trim operation must delete all original data in invalid and valid blocks throughout the device. To perform a full trim operation, a comprehensive analysis of the history of garbage collection and deletion is required. This places a significant burden on the management of storage devices.

To address these inherent challenges, the on-chip secure deletion scheme should be applied in real-time. Regardless of trim requests, the storage device can apply the on-chip secure deletion scheme in real-time when performing data updates or deletions. This prevents the existence of original data in invalidated pages, which may occur naturally. Moreover, it eliminates the fundamental existence of original data in the invalidation block. By implementing the on-chip secure deletion scheme, the original data exists only in the validated pages of the validated block, significantly simplifying the management of original data.

3.2. Performance comparison of Off-chip/On-chip schemes

The comparison between the off-chip secure deletion scheme and the on-chip secure deletion scheme is as follows. The off-chip secure deletion scheme primarily consists of an encryption scheme and a trim scheme that includes an erase operation. On the other hand, the on-chip secure deletion scheme encompasses a scrubbing scheme [21], a partial overwriting scheme [12], a down-bit programming scheme [22, 23], a deletion pulse application scheme [23, 25-27], and the proposed code modulation secure deletion scheme. The performance of the off-chip secure deletion scheme and the on-chip secure deletion scheme is compared in Table I as follows:

TABLE I: Performance Differences in Off-chip/On-chip Secure Deletion

| Approach | Deletion Size | Real-time Response | Overhead |
| --- | --- | --- | --- |
| Off-chip Secure Deletion | Block Unit | Impossible | Management Cost |
| On-chip Secure Deletion | Page Unit | Possible | Program Disturbance |

In the off-chip secure deletion scheme, the erase operation is performed in units of blocks. Typically, NAND flash memory devices do not frequently execute erase operations at the block level due to management concerns. It is commonly understood that NAND flash memory manufacturers guarantee 1000 erase operations per block. If a block undergoes more than 1000 erase operations, the usability of the storage device becomes compromised. Consequently, the off-chip secure deletion scheme cannot achieve a real-time trim operation. The absence of real-time trim operation results in a cost associated with block management. On the other hand, the on-chip secure deletion scheme enables erase operations to be performed in units of pages.

3.3. Traditional On-chip secure deletion schemes

Existing on-chip secure deletion schemes primarily focus on modulation with the original data using overwriting technology. The scrubbing scheme transforms the threshold voltage of a memory cell with the most significant bit, while the partial overwriting scheme generates and reprograms possibly random data. The deletion pulse application scheme applies a pulse to modulate the original data to a point where error correction becomes impossible in ECC. The scrubbing scheme requires a significant amount of time to program memory cells up to the most significant bit. Moreover, excessive execution of high-level state programming can lead to physical destruction of the page. While the partial overwriting scheme reduces the possibility of physical destruction compared to the scrubbing scheme, it adds the time required to generate and program programmable random data. In comparison, the deletion pulse application scheme can efficiently and simply delete original data at a page level while reducing the risk of physical destruction. However, there is a concern regarding the possibility of data modification connected to an adjacent valid wordline, which can cause erase pulse disturbance.

4. Proposed On-Chip Secure Deletion Scheme

For a complete trim operation by the host, on-chip secure deletion becomes an essential technology that must be applied. Depending on the host or customer's requirements, on-chip secure deletion is likely to develop into an essential technology for storage devices.

4.1. Basic ECC operation

Data stored in NAND flash memory is prone to leakage current and data deformation due to program/read disturbance. To ensure data reliability, various technologies have been developed, including the application of error correction schemes for data recovery. Parity, necessary for recovering the original programmed data, is generated, and during a write operation, the original data and parity are simultaneously stored. During a read operation, the original data and parity are read, and errors in the original data are corrected using the parity. Referring to Fig. 1, NAND flash memory devices consist of multiple planes, each containing blocks connected to wordlines and bitlines. Pages corresponding to wordlines are present within each block, and a spare area is available for error management functions.

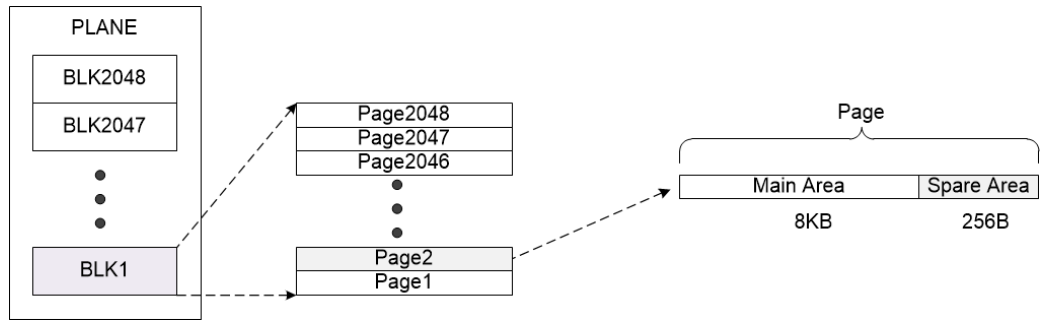

Figure 1. Physical page configuration of a general NAND Flash Memory. The physical page includes a main area for storing user data and a spare area for storing meta data. Here, the metadata includes ECC of user data.

The ECC engine includes a parity generator, as shown in Fig. 2. This generator produces parity data using the input data of a target page in the NAND flash memory device. Parity data can be generated by segments of the input data (1 KB out of 4 KB) or by utilizing the entire input data. The input data is programmed into the main area of the target page, while the parity data is programmed into the spare area of the target page.

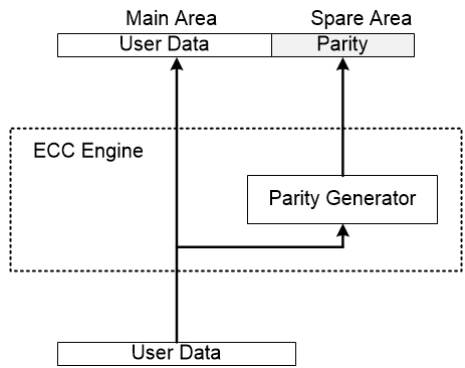

Figure 2. ECC encoding process. ECC engine generates parity corresponding to user data.

The ECC engine consists of a syndrome generator, Berlekamp block, Chien block, and data corrector, as depicted in Fig. 3. The syndrome generator calculates or generates syndromes using the output data and parity data read from the NAND flash memory device to determine if an error exists. Syndromes are then fed into an error locator polynomial and a Berlekamp block, which determines the number of errors. The Chien block finds the square roots of the polynomial in the error locator polynomial output by the Berlekamp block. Finally, if there are any errors based on the output of the Chien block, the data corrector corrects the output data, resulting in corrected data.

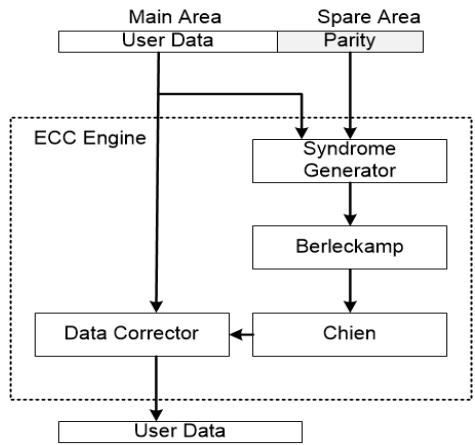

Figure 3. ECC Decoding Process.

### 4.2. Parity modulation-based secure deletion

We propose a novel on-chip secure deletion scheme that primarily utilizes parity modulation. Modulating a relatively small parity instead of a relatively large original data is likely to reduce the time and cost required for secure deletion. Therefore, we suggest a parity modulation-based secure deletion scheme when performing secure deletion based on the host's request, as shown in Fig. 4. For this purpose, a partial program operation may be performed on the data of the spare area connected to the wordline related to the original data. The partial program operation is generally defined not to exceed four or more times for one wordline.

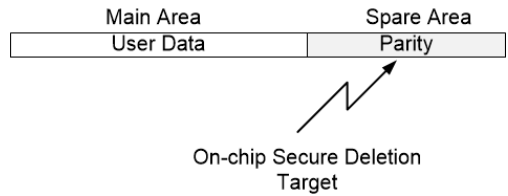

Figure 4. On-Chip Secure Deletion Target.

### 4.3. Partial programming for spare area

The proposed on-chip secure deletion scheme includes a partial program operation on the spare area where parity is stored, as depicted in Fig. 5. In the partial program operation, the program can be performed with zero bits or a predetermined program pulse can be applied to the part where the parity is stored. It is important for NAND flash memory to support "small data move" in its basic functionality.

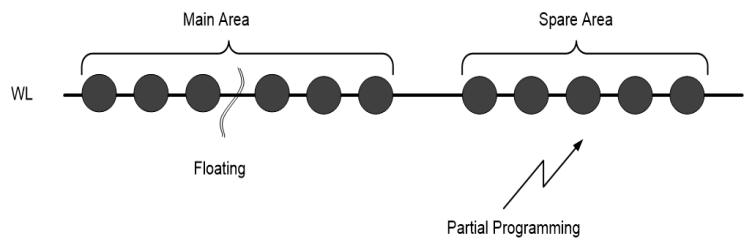

Figure 5. Partial programming for spare area.

### 4.4. Error count increase and read-fail output

The ECC engine has a specific number of error-correctable bits, which gradually increases with technological advancements. When error correction is impossible, a NAND flash memory device typically performs a recovery code. The recovery code conducts a page read operation in various ways. If error correction is still impossible after the recovery code, the NAND flash memory device eventually transmits a read fail signal to the host. In the proposed method, there is a high possibility that error correction becomes impossible even when applying the defense code through modulation of the parity data. Consequently, a read fail signal is outputted to the host, as shown in Fig. 6. Unlike existing data modulation schemes, the proposed scheme does not require additional verification operations to prove the deletion of original data. It can be immediately confirmed through the on-chip secure deletion operation that the read operation for invalid/valid pages is impossible.

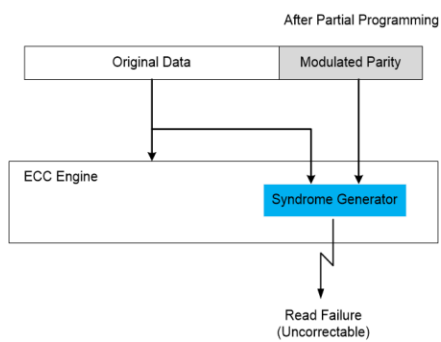

Figure 6. Real-time read failure output.

## 4.5. Secure storage device and operating method

The storage device we propose can be referred to as a secure storage device because it performs secure deletion. The secure storage device, as shown in Fig. 7, includes a NAND flash memory device and a controller (CTRL) that controls it. The controller (CTRL) receives a TRIM command from an external host device and performs a Secure Deletion operation accordingly. This secure deletion operation is performed in conjunction with an ECC circuit. When the TRIM command is received, the controller starts the secure deletion operation in real time.

The ECC circuit can generate ECC data (ECC_SD) to be stored in the physical space where secure deletion is required, for example, in the spare area. Where ECC data (ECC_SD) is generated as a value that directs user data to be uncorrectable by errors. For this, a process of reading data on the page to be deleted can precede. On the other hand, ECC data (ECC_SD) can also be set to a fixed value in accordance with a secure deletion request. This fixed value can be determined experimentally.

At the same time, the controller can generate user data stored in the main area. The user data used in the secure deletion operation can be composed of data in the lowest state (for example, erase state). This is because the program operation of the lowest state effectively corresponds to the prohibition of the program. In other words, the proposed secure deletion operation can be expected to minimize the program disturbance of neighboring pages by tampering with error correction codes and programming the lowest state data.

Generally, the memory cells of a NAND flash memory device have multiple states. Among these states, the lowest state (for instance, the erase state) basically becomes a program pass without applying a program pulse due to its lowest threshold voltage. Therefore, when programming with the lowest state data, it can be understood to have a practical program prohibition effect. Also, if partial programming operation is possible, only the part including the ECC area can be processed for secure deletion operation with the new error correction code (ECC_SD).

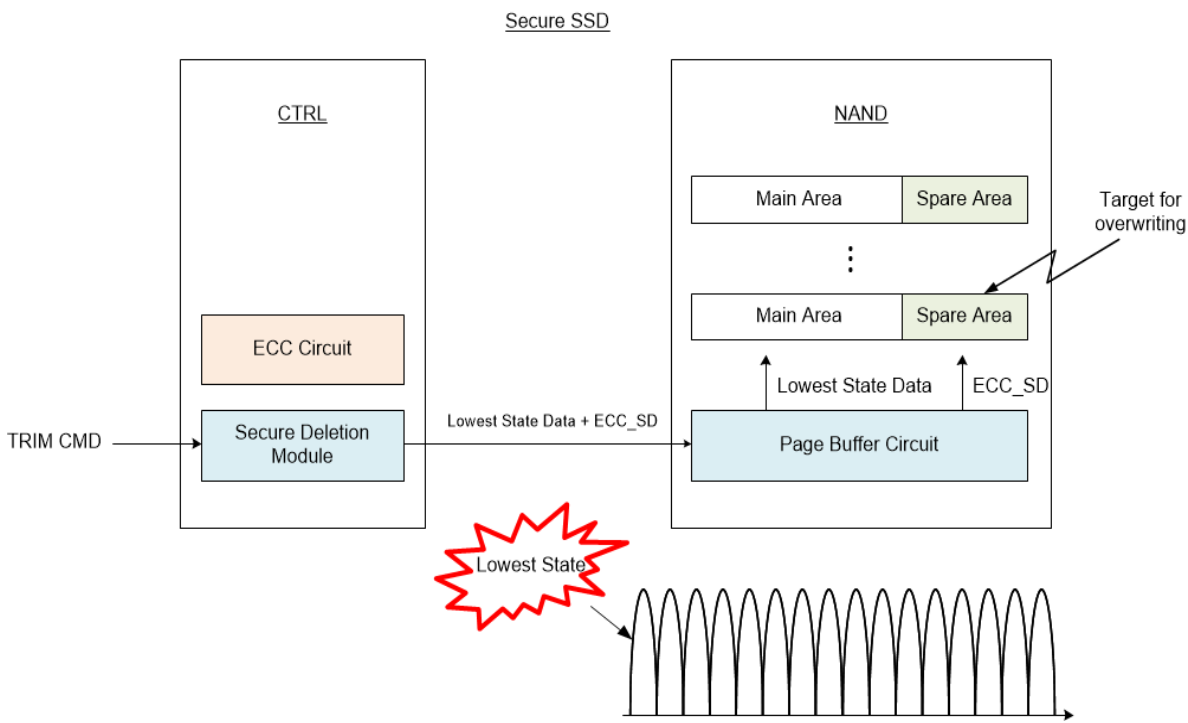

Figure 7. Secure Solid State Device. The secure storage device executes the secure deletion module in response to the trim command. The secure deletion module generates user data configured in the lowest state and a modified ECC value, and performs a program operation on the page to be deleted with the generated page data.

When the host transfers a secure deletion request for a file stored on the storage device SSD, as illustrated in Fig. 8, the controller of the storage device translates logical addresses associated with the corresponding file into physical addresses in response to the host's secure deletion request. The controller generates ECC parity for secure deletion and transmits a command for performing a partial program operation on the ECC area to the NAND flash memory device through wordlines corresponding to physical addresses. The NAND flash memory device performs a partial program operation using ECC parity generated for the spare area in response to the partial program command. The non-volatile memory device then transmits a completion message for the partial program operation to the controller. Subsequently, the controller sends a read command to the non-volatile memory device through the corresponding wordline, and the non-volatile memory device responds by transmitting the read data to the controller. The controller determines whether error correction for the read data is possible. If error correction is unachievable, the controller sends a secure deletion completion message to the host. The described secure deletion operation can be performed in conjunction with the defense code (or read retry) of the storage device. Alternatively, the storage device may delete the separate ECC parity generated for secure deletion and apply a predetermined number of erase pulses to the spare area storing ECC.

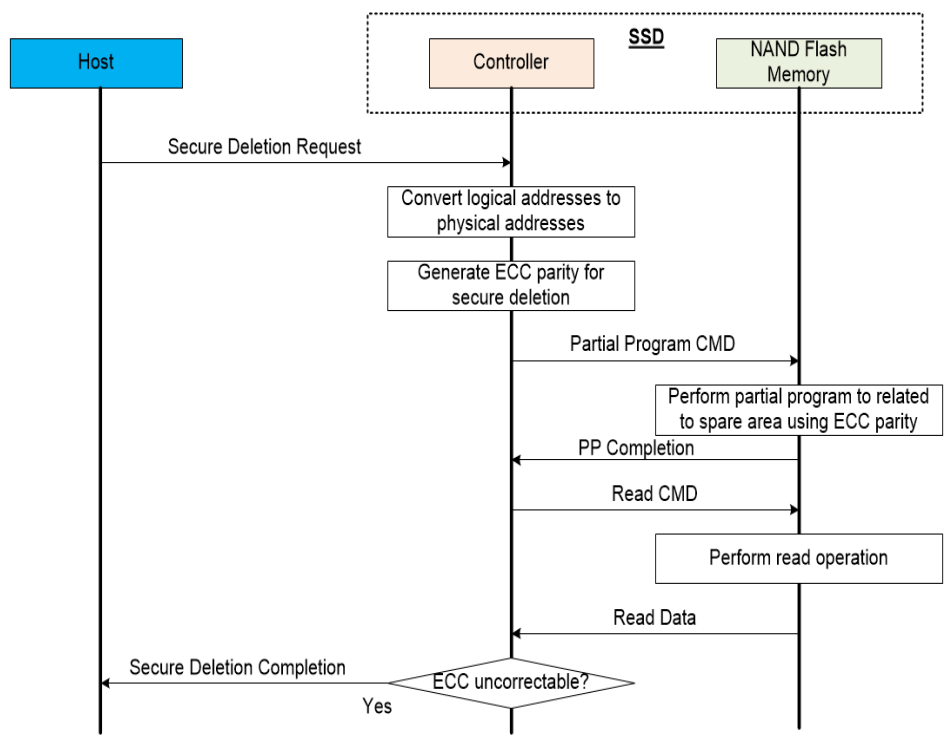

Figure 8. Total process for proposed secure deletion using ECC parity. The SSD includes a controller and a NAND flash memory. The controller performs ECC modulation using partial programming in response to the secure erase request, and transmits a secure erase complete message to the host when error correction is impossible.

The proposed secure deletion technique allows for real-time deletion and verification. Furthermore, since the proposed secure deletion technique does not involve program operations on data, it significantly reduces disturbances on valid data compared to existing on-chip secure deletion techniques. It is worth exploring the implementation of the ECC engine in an on-chip structure within NAND flash memory, as this concept holds promise for future research on secure deletion.

5. Results and Analysis

The scrubbing scheme, partial overwriting scheme, down programming scheme, deletion pulse application scheme, and proposed code modulation scheme can all delete original data at the wordline level. Since these schemes operate at the page level, real-time trim operations are possible. However, the page-level scheme applies a predetermined voltage level to the wordline, which can have an adverse effect on adjacent valid pages. This disadvantage manifests as program disturbance. In the following section, we compare the performance of on-chip secure deletion schemes, as indicated in Table II.

TABLE II: Performance Differences of On-chip Secure Deletion Schemes

| On-chip Secure Deletion Scheme | Data Generation | Endurance | Program Disturbance | Verification Reporting |
|---|---|---|---|---|
| Scrubbing [10], [20] | Zero Bit | Increase in cell wear | High | None |
| Partial Overwriting [12] | Possible Random Bit | None | Medium | None |
| Down-Bit Programming [21],[22],[23] | SLC data bit | None | Low | None |
| Deletion Pulse Application [23],[25],[26],[27] | None | None | Low | None |
| Code Modulated Secure Deletion (Proposed Scheme) | In some cases | None | Very Small | Read Failure |

Basically, on-chip deletion schemes rely on overwriting, which requires the generation of random data. In the scrubbing scheme [10, 20], the most significant bit, specifically the zero bit, is generated. The partial overwriting scheme [12] involves generating arbitrary data in a programmable state relative to the programmed data in multi-level cells. The down-bit programming scheme [21-23] requires data conversion from multi-level cells to single-level cells. In this case, separate management is necessary to ensure that the storage capacity remains unchanged. The deletion pulse application scheme [23, 25-27] applies multiple deletion pulses instead of programming standardized data. This scheme eliminates the need for data creation time compared to other on-chip secure deletion schemes. The code modulation secure deletion scheme may also generate arbitrary data in certain cases. It can be implemented by applying multiple pulses similar to the deletion pulse application scheme.

Next, the durability-related indices are compared as follows. The scrubbing scheme, due to programming in the highest state, significantly increases the wear rate of the cell. In terms of durability, it may be more favorable to perform erase operations at the block level rather than frequently using the scrubbing scheme. The partial overwriting scheme, down-bit programming scheme, deletion pulse application scheme, and code modulation secure deletion scheme do not have durability issues. However, since on-chip secure deletion schemes are primarily based on overwriting, they inevitably cause program disturbance. Program disturbance generally refers to the degree of data destruction in memory cells connected to wordlines adjacent to the wordline receiving the program pulse [37]. Therefore, if the retrieval and time of program pulse application in secure deletion operations are relatively high, the program disturbance is higher compared to cases where they are not. We have roughly categorized the degree of program disturbance into high, medium, low and very small. This indicates the approximate number of program pulses that need to be applied until the secure deletion operation is completed. The scrubbing scheme, which writes data in the highest state, necessitates proceeding to a higher program level, resulting in relatively high program disturbance. Data management of adjacent valid pages is likely to be required. Similar to the scrubbing scheme, the partial overwriting scheme may cause program disturbance, albeit to a lesser extent as it does not involve the highest state data like the scrubbing scheme. The down-bit programming scheme, deletion pulse application scheme, and code modulation secure deletion scheme cause relatively very small program disturbance compared to the scrubbing, partial overwriting scheme, and deletion pulse application. The proposed ECC modulation technique adopts programming user data to the lowest state of data. The lowest state of data is effectively a program inhibit as it mostly passes through the program path in a single program pulse. The proposed technique requires only the ECC modulation program to pass, resulting in a relatively low number of program pulse applications compared to other secure deletion techniques until it is relatively complete. In other words, the number of program pulses applied to the selected wordline until the proposed ECC modulation program succeeds is lower compared to other secure deletion techniques. Therefore, the program disturbance due to secure deletion is very low compared to that of other techniques

Finally, in the scrubbing scheme, partial overwriting scheme, down-bit programming scheme, and deletion pulse application scheme, a separate operation is required to verify whether the original data has been completely deleted in the NAND flash memory device. However, in the proposed code-modulated secure deletion scheme, a read failure is produced during a read operation, eliminating the need for a separate verification operation. This is because the ECC values have been altered according to the secure deletion technique. Even if the original data still exists in the user area, attempting to perform a read operation based on the associated physical address will result in a read failure. The storage device can only output information stating that it is unable to read the corresponding page of the NAND flash memory. This serves as a real-time indication of the completion of secure deletion. Such a technique can be implemented cost-effectively and facilitates easy secure deletion without altering the basic operations of existing NAND flash memory.

6. Conclusion

　We have examined the limitations of the trim operation in secure deletion. Existing off-chip secure deletion schemes are primarily based on an erase operation, which prevents them from performing a real-time trim operation in terms of management. On the contrary, on-chip secure deletion schemes allow for a real-time trim operation as they operate on a page-by-page deletion basis. These on-chip secure deletion schemes ensure a complete trim operation compared to off-chip secure deletion schemes. However, the on-chip secure deletion schemes may cause program disturbance due to the application of pulses to wordlines at the page level. In contrast, the proposed on-chip secure deletion scheme employs ECC modulation through partial programming to simplify and secure deletion while minimizing the program disturbance problem. The ECC modulation secure deletion scheme minimizes the program disturbance problem and allows for real-time verification of the deleted original data compared to other on-chip secure deletion schemes. This proposed secure deletion scheme can be effectively applied to verify data in the invalidated area of the storage device. In the future, the code modulation secure deletion scheme, proposed as an anti-forensic technology for storage devices, is highly likely to be implemented.

7. References

[1] Z. Istvan, S. Ponnapalli and V. Chidambaram, "Software-defined data protection: low overhead policy compliance at the storage layer is within reach!", Proceedings of the VLDB Endowment, vol. 14, no. 7, 2021.

[2] A. Tosa, A. Hangan, G. Sebestyen and Z. István, "In-Storage Computation of Histograms with differential privacy," *2021 International Conference on Field-Programmable Technology (ICFPT)*, 2021, pp. 1-4, doi: 10.1109/ICFPT52863.2021.9609899.

[3] Y. Wang, C. -F. Chen, P. -Y. Kong, H. Li and Q. Wen, "A Cyber–Physical–Social Perspective on Future Smart Distribution Systems," in *Proceedings of the IEEE*, 2022, doi: 10.1109/JPROC.2022.3192535.

[4] R. A. Grimes. "Cryptography Apocalypse: Preparing for the Day When Quantum Computing Breaks Today's Crypto," John Wiley & Sons, 2019.

[5] K. Denker and A. Y Javaid, "Quantum Computing as a Threat to Modern Cryptography Techniques", Proceedings of the International Conference on FCS (Foundations of Computer Science), pp. 3-8, 2019.

[6] K. Sun, J. Choi, D. Lee and S. H. Noh, "Secure Deletion of Confidential Data in Consumer Electronics," *2008 Digest of Technical Papers - International Conference on Consumer Electronics*, 2008, pp. 1-2, doi: 10.1109/ICCE.2008.4588029.

[7] D. W. Byunghee Lee, Kyungho Son and S. Kim, "Secure data deletion for usb flash memory," Journal of Information Science and Engineering, pp. 1710-1714, 2011.

[8] S. Jia, L. Xia, B. Chen and P. Liu, "NFPS: Adding undetectable secure deletion to flash translation layer", Proc. 11th ACM Asia Conf. Comput. Commun. Security (ASIA CCS), pp. 305-316, 2016.

[9] Q. Zhang et al., "Ensuring Data Confidentiality with a Secure XTS-AES Design in Flash Translation Layer," 2020 IEEE 5th International Conference on Cloud Computing and Big Data Analytics (ICCCBDA), 2020, pp. 289-294, doi: 10.1109/ICCCBDA49378.2020.9095700.

[10] J. Cui, W. Liu, J. Huang and L. T. Yang, "ADS: Leveraging Approximate Data for Efficient Data Sanitization in SSDs," in *IEEE* Transactions on Computer-Aided Design of Integrated Circuits an*d Systems*, vol. 41, no. 6, pp. 1771-1784, June 2022, doi: 10.1109/TCAD.2021.3100274.

[11] M. Kim et al., "Evanesco: Architectural support for efficient data sanitization in modern flash-based storage systems", Proc. Int. Conf. Archit. Support Program. Lang. Oper. Syst., pp. 1-16, 2020.

[12] N. Y. Ahn and D. H. Lee, "Duty to delete on non-volatile memory", arXiv:1707.02842, Jul. 2017, [online] Available: https://arxiv.org/abs/1707.02842.

[13] S. Diesburg, C. Meyers, M. Stanovich, A.-I. A. Wang and G. Kuenning, "TrueErase: Leveraging an auxiliary data path for per-file secure deletion", ACM Trans. Storage, vol. 12, no. 4, pp. 18, 2016.

[14] J. Kwak, H. C. Kim, I. H. Park and Y. H. Song, "Anti-forensic deletion scheme for flash storage systems," 2016 IEEE International Conference on Network Infrastructure and Digital Content (IC-NIDC), 2016, pp. 317-321, doi: 10.1109/ICNIDC.2016.7974588.

[15] B. Chen, S. Jia, L. Xia and P. Liu, "Sanitizing data is not enough! Towards sanitizing structural artifacts in flash media", Proc. 32nd Annu. Conf. Comput. Security Appl. (ACSAC), pp. 496-507, 2016.

[16] L. Yang, T. Wei, F. Zhang and J. Ma, "SADUS: Secure data deletion in user space for mobile devices", Comput. Security, vol. 77, pp. 612-626, Aug. 2018.

[17] T. Ojo, H. Chi, J. Elliston and K. Roy, "Secondhand Smart IoT Devices Data Recovery and Digital Investigation," SoutheastCon 2022, 2022, pp. 640-648, doi: 10.1109/SoutheastCon48659.2022.9763996.

[18] T. A, "An Empirical Comparison of Smartphone Factory-Resets to Remote Deletion Applications," University of South Alabama ProQuest Dissertations Publishing, 2019. 13857756.

[19] J. A. Yaacoub, et. al, "Advanced digital forensics and anti-digital forensics for IoT systems: Techniques, limitations and recommendations," Internet of Things, Vol. 19, 2022, https://doi.org/10.1016/j.iot.2022.100544.

[20] W. -C. Wang, C. -C. Ho, Y. -H. Chang, T. -W. Kuo and P. -H. Lin, "Scrubbing-Aware Secure Deletion for 3-D NAND Flash," in *IEEE Transactions on Computer-Aided Design of Integrated Circuits and Systems*, vol. 37, no. 11, pp. 2790-2801, Nov. 2018, doi: 10.1109/TCAD.2018.2857260.

[21] P. -H. Lin, Y. -M. Chang, Y. -C. Li, W. -C. Wang, C. -C. Ho and Y. -H. Chang, "Achieving Fast Sanitization with Zero Live Data Copy for MLC Flash Memory," *2018 IEEE/ACM International Conference on Computer-Aided Design (ICCAD)*, 2018, pp. 1-8, doi: 10.1145/3240765.3240773.

[22] B. Li and D. H. C. Du, "TASecure: Temperature-Aware Secure Deletion Scheme for Solid State Drives", *Proceedings of the 2019 on Great Lakes Symposium on VLSI*, pp. 275-278, May 2019.

[23] N. Y. Ahn and D. H. Lee, "Schemes for Privacy Data Destruction in NAND Flash Memory," in IEEE Access, vol. 7, pp. 181305-181313, 2019, doi: 10.1109/ACCESS.2019.2958628.

[24] W. -C. Wang, C. -C. Ho, Y. -M. Chang and Y. -H. Chang, "Challenges and Designs for Secure Deletion in Storage Systems," 2020 Indo – Taiwan 2nd International Conference on Computing, Analytics and Networks (Indo-Taiwan ICAN), 2020, pp. 181-189, doi: 10.1109/Indo-TaiwanICAN48429.2020.9181335.

[25] N. Y. Ahn and D. H. Lee, "Forensics and Anti-Forensics of NAND Flash Memory: From a Copy-Back Program Perspective," in IEEE Access, vol. 9, pp. 14130-14137, 2021, doi: 10.1109/ACCESS.2021.3052353.

[26] N. Y. Ahn and D. H. Lee, "Forensic Issues and Techniques to Improve Security in SSD with Flex Capacity Feature," in IEEE Access, vol. 9, pp. 15130-15137, 2021, doi: 10.1109/ACCESS.2021.3136483.

[27] N. Y. Ahn and D. H. Lee, "Security of IoT Device: Perspective Forensic/Anti-Forensic Issues on Invalid Area of NAND Flash Memory," in *IEEE Access*, vol. 10, pp. 74207-74219, 2022, doi: 10.1109/ACCESS.2022.3190957.

[28] M. Wang, Q. Zhang, "Optimized data storage algorithm of IoT based on cloud computing in distributed system," Computer Communications, Vol. 157, pp. 124-131, 2020, https://doi.org/10.1016/j.comcom.2020.04.023.

[29] R. Subramani, H. Swapnil, N. Thakur, B. Radhakrishnan and K. Puttaiah, "Garbage Collection Algorithms for NAND Flash Memory Devices -- An Overview," *2013 European Modelling Symposium*, 2013, pp. 81-86, doi: 10.1109/EMS.2013.14.

[30] R. Jin, H. -j. Cho and T. -s. Chung, "An encryption approach to secure modification and deletion for flash-based storage," in *IEEE Transactions on Consumer Electronics*, vol. 60, no. 4, pp. 662-667, Nov. 2014, doi: 10.1109/TCE.2014.7027340.

[31] B. Park, "A Novel Recovery Data Technique on MLC NAND Flash Memory," *2019 11th International Conference on Knowledge and Systems Engineering (KSE)*, 2019, pp. 1-5, doi: 10.1109/KSE.2019.8919382.

[32] B. Fitzgerald, C. Ryan and J. Sullivan, "An Early-Life NAND Flash Endurance Prediction System," in *IEEE Access*, vol. 9, pp. 148635-148649, 2021, doi: 10.1109/ACCESS.2021.3124604.

[33] Y. -M. Lin, H. -T. Li, M. -H. Chung and A. -Y. Wu, "Byte-Reconfigurable LDPC Codec Design With Application to High-Performance ECC of NAND Flash Memory Systems," in *IEEE Transactions on Circuits and Systems I: Regular Papers*, vol. 62, no. 7, pp. 1794-1804, July 2015, doi: 10.1109/TCSI.2015.2423798.

[34] J. Li et.al, "Intra-page Cache Update in SLC-mode with Partial Programming in High Density SSDs," ICPP 2021: 50th International Conference on Parallel Processing, No. 46, pp. 1–10, 2021, https://doi.org/10.1145/3472456.3472492

[35] H. Chen, Z. Sun, F. Yi and J. Su, "Bufferbank storage: An economic scalable and universally usable in-network storage model for streaming data applications", *Sci. China Inf. Sci.*, vol. 59, no. 1, pp. 1-15, 2016.

[36] P. Kumari, U. Surendranathan, M. Wasiolek, K. Hattar, N. P. Bhat and B. Ray, "Radiation-Induced Error Mitigation by Read-Retry Technique for MLC 3-D NAND Flash Memory," in *IEEE Transactions on Nuclear Science*, vol. 68, no. 5, pp. 1032-1039, May 2021, doi: 10.1109/TNS.2021.3052909.

[37] X. Jia et al., "A Novel Program Scheme to Optimize Program Disturbance in Dual-Deck 3D NAND Flash Memory," in IEEE Electron Device Letters, vol. 43, no. 7, pp. 1033-1036, July 2022, doi: 10.1109/LED.2022.3178155.